\documentclass[draft]{agujournal2019}
\usepackage{url} %this package should fix any errors with URLs in refs.
\usepackage[inline]{trackchanges} %for better track changes. finalnew option will compile document with changes incorporated.
\usepackage{soul}
\usepackage{amsmath}
\usepackage{natbib}
\draftfalse

\journalname{Space Weather}

\begin{document}

% (A title should be specific, informative, and brief. Use
% abbreviations only if they are defined in the abstract. Titles that
% start with general keywords then specific terms are optimized in
% searches)

\nolinenumbers

%\title{Differential Orbital Decay of the Van Allen Probes Under Thermospheric Drag}

\title{What Caused the Orbital Decay of the Twin Van Allen Probes to Diverge?} %Role of Thermospheric Drag and Space Weather}

% The Curious Case of the Van Allen Probes: Differential Orbital Decay Explains an Unexpected Reentry Sequence

\authors{Ayisha M. Ashruf\affil{1, 2}, Ankush Bhaskar\affil{1}, C. Vineeth\affil{1}, Tarun Kumar Pant\affil{1}, Anand Narayanan\affil{2}}

\affiliation{1}{Space Physics Laboratory, Vikram Sarabhai Space Centre, Thiruvananthapuram, Kerala, India}
\affiliation{2}{Indian Institute of Space Science and Technology, Thiruvananthapuram, Kerala,	India}
\correspondingauthor{Ayisha M Ashruf}{ayishamashruf@gmail.com}

%%%%%%%%%%%%%%%%%%%%%%%%%%%%%%%%%%%%%%%%%%%%%%%
% KEY POINTS
%%%%%%%%%%%%%%%%%%%%%%%%%%%%%%%%%%%%%%%%%%%%%%%
%  List up to three key points (at least one is required)
%  Key Points summarize the main points and conclusions of the article
%  Each must be 140 characters or fewer with no special characters or punctuation and must be complete sentences

% Example:
% \begin{keypoints}
% \item	List up to three key points (at least one is required)
% \item	Key Points summarize the main points and conclusions of the article
% \item	Each must be 140 characters or fewer with no special characters or punctuation and must be complete sentences
% \end{keypoints}

\begin{keypoints}
\item Small orbital differences evolved into a runaway drag feedback that accelerated the decay of Van Allen Probe A
\item Geomagnetic storms amplified orbital decay differences between two spacecraft exposed to the same space weather conditions
\item Space weather can produce divergent satellite lifetimes even when spacecraft begin with slight differences in orbits
\end{keypoints}

\begin{abstract}
The Van Allen Probes were launched into nearly identical orbits and experienced the same solar and geomagnetic environment, yet their post-mission lifetimes diverged dramatically. Probe A underwent rapidly accelerating orbital decay and reentered the atmosphere in March 2026, while Probe B remains in orbit. We investigate the cause of this divergence using reconstructed orbital trajectories, modeled thermospheric densities, and measures of low-altitude residence and atmospheric drag. We find that small differences in orbital evolution progressively increased Probe A’s residence time in the lower thermosphere, where the steep density gradient made additional low-altitude exposure increasingly important. Enhanced thermospheric density during periods of geomagnetic activity further amplified this difference, leading to greater drag and accelerating orbital decay of Probe A. Our results show how small differences in orbital evolution can be amplified through the coupled effects of thermospheric drag and space weather, producing substantially different orbital lifetimes even for twin spacecraft sharing nearly identical initial orbits.
\end{abstract}

\section*{Plain Language Summary}
The Van Allen Probes were launched together and initially followed very similar orbits, but a few years after the end of the mission, their orbital decay began to diverge. Van Allen Probe A gradually spent more time at lower altitudes, where the atmosphere is much denser. This exposed Probe A to greater atmospheric drag, causing it to lose orbital energy faster and descend even further. Geomagnetic storms and increasing solar activity likely amplified this difference by temporarily increasing thermospheric density. The results suggest that small differences in orbital evolution can grow through a feedback process, ultimately leading to substantially different orbital lifetimes, even for spacecraft that began with nearly identical orbits and experienced the same space weather conditions. Probe A consequently reentered in March 2026, while Probe B remained in orbit.
\section{Introduction}

Orbital decay of satellites in Low-Earth Orbit (LEO) is governed primarily by atmospheric drag, which depends on both the density of the upper atmosphere and the spacecraft's orbital characteristics. Variations in thermospheric density driven by solar and geomagnetic activity can substantially modify satellite lifetimes, particularly for spacecraft in highly elliptical orbits that repeatedly traverse the dense lower thermosphere \citep{EMMERT2015ThermMassDen, FULLERROWELL2018ThermIonReactionSpaceWeather, HUFFMAN1992UpperAtmCompnDen}. Moreover, extreme geomagnetic storms can significantly accelerate the decay of space objects \citep{tsurutani2003extreme, hayakawa2020extreme,thampi2025extreme}. Therefore, accurate prediction of orbital decay is essential for mission operations, space situational awareness, and long-term assessments of the orbital debris environment.

Although atmospheric drag has been studied extensively, predicting the long-term orbital evolution of individual spacecraft remains challenging because the thermospheric density exhibits significant variability over a wide range of temporal and spatial scales. In addition to changes associated with the solar cycle \citep{Ayisha2026}, geomagnetic disturbances can produce substantial variations in the drag experienced by spacecraft following similar orbital trajectories \citep{Oliveira2019SatDragMagStorms,oliveira2020estimating,baruah2024loss}. Thus, small differences in orbital geometry can accumulate over time, resulting in markedly different orbital evolution.

The Van Allen Probes provide a unique opportunity to investigate these processes. Launched together in August 2012 into identical, highly elliptical Earth orbits, the two spacecraft, Van Allen Probe A and Van Allen Probe B, were designed to operate as a coordinated mission to study the dynamics of Earth's radiation belts. Both spacecraft were placed in orbits inclined by 10.2$^\circ$ and shared the same octagonal-cylinder configuration with four solar panels and eight rods. Following the successful completion of their primary science mission, both spacecraft remained in orbit, where their trajectories gradually decayed under the influence of atmospheric drag. Van Allen Probe B was deactivated on 19 July 2019, followed by Van Allen Probe A on 18 October 2019. Based on pre-mission lifetime projections, both spacecraft were expected to remain in orbit until the year 2034. However, contrary to these predictions, Van Allen Probe A reentered Earth's atmosphere in March 2026, while Van Allen Probe B remained in orbit despite having been decommissioned several months earlier. This unexpected divergence raises an important question: how did two spacecraft with identical initial orbital configurations evolve to exhibit substantially different orbital lifetimes?

In this study, we investigate the physical mechanisms responsible for the divergent orbital evolution of the Van Allen Probes during their post-mission decay phase. Rather than focusing solely on instantaneous orbital parameters such as perigee altitude, we quantify the cumulative atmospheric interaction experienced by each spacecraft using integrated measures of thermospheric drag exposure and orbital energy loss. These diagnostics are combined with analyses of orbital geometry and atmospheric density sampling to determine what caused the accelerated decay of Van Allen Probe A.

\section{Data and Methodology}
\subsection{Data}

The post-deactivation orbital evolution of Van Allen Probes A and B was examined using publicly available two-line element (TLE) data obtained from the Space-Track website. The analysis extends from the deactivation of each spacecraft through the atmospheric reentry of Van Allen Probe A on 11 March 2026. TLEs provide mean orbital elements for use with the SGP4 propagator, and the resulting trajectories represent reconstructed orbital states rather than actual position measurements. Because propagation errors increase with the time elapsed from a TLE epoch, each TLE was propagated only over the interval to the subsequent available TLE, thereby limiting the extrapolation time.

The TLEs were propagated using the `EarthSatellite` module of the Skyfield Python library at 15-second intervals. The resulting state vectors were used to obtain spacecraft position and velocity, from which altitude, latitude, longitude, and orbital parameters were determined. The 15-second sampling provides sufficient temporal resolution to resolve the short-duration perigee passages and to evaluate the atmospheric density integral over individual revolutions.

Atmospheric neutral density along the satellite orbit was computed using the NRLMSIS 2.0 empirical atmospheric model \citep{Emmert2020MSIS}. The model uses the geographic coordinates and time together with solar and geomagnetic activity indices to estimate the total neutral mass density. For each propagated spacecraft position, the model was supplied with the corresponding altitude, latitude, longitude, time, F10.7, and A$_\mathrm{p}$ indices. The resulting density was assigned to the corresponding spacecraft position and used in the subsequent analysis. For each spacecraft, time series of semi-major axis, latitude, longitude, and atmospheric density were generated throughout the post-deactivation period. 

\subsection{Orbital Evolution}

To characterize the long-term orbital evolution, the temporal variations of the semi-major axis and perigee altitudes were examined for both spacecraft. The residence time within the lower thermosphere was quantified by calculating, for each orbital revolution, the total duration for which the spacecraft altitude remained below specific altitudes. This measure complements perigee altitude by quantifying the duration of atmospheric exposure rather than only the minimum altitude reached during a revolution. The residence time was calculated over a continuous range of critical altitudes between 200 and 700 km to determine the altitude at which the cumulative difference between the two spacecraft is largest. To examine whether differences in orbital geometry could lead to different thermospheric sampling under common external forcing, the geographic latitude, longitude, and local solar time at each perigee passage were also determined. These parameters characterize the location and local-time sector of the thermosphere sampled near the point of maximum atmospheric drag. 

\subsection{Integrated Atmospheric Drag}

Atmospheric drag accumulated over an orbit depends on both the atmospheric density and the duration and velocity of the spacecraft's passage through the lower thermosphere. To quantify this cumulative interaction independently of the spacecraft-specific ballistic coefficient, an integrated drag exposure proxy was calculated for each orbital revolution,

\begin{equation}
	D_{\rho v^2} = \int_{\mathrm{orbit}} \frac{1}{2} \rho v_{\mathrm{rel}}^2\,dt,
\end{equation}

where $\rho$ is the atmospheric mass density and $v_{\mathrm{rel}}$ is the spacecraft velocity relative to the surrounding atmosphere. Since the magnitude of atmospheric velocity is much smaller compared to spacecraft velocity ($v$), we assume $v_{\mathrm{rel}} \sim v$. The corresponding physical drag impulse is obtained from

\begin{equation}
	I_D = \frac{1}{2}C_D A D_{\rho v^2},
\end{equation}

where $C_D$ is the drag coefficient and $A$ is the effective velocity-facing area.

An analogous integrated energy-loss proxy was calculated as

\begin{equation}
	D_{\rho v^3} = \int_{\mathrm{orbit}} \frac{1}{2} \rho v^3\,dt.
\end{equation}

The corresponding specific orbital-energy loss due to drag is

\begin{equation}
	\Delta E_{\mathrm{drag}} = -\frac{1}{2}\frac{C_DA}{m} D_{\rho v^3},
\end{equation}

where $m$ is the spacecraft mass. 

The two integrated quantities therefore provide complementary measures of cumulative atmospheric interaction and drag-induced orbital energy dissipation. Because both spacecraft have similar physical properties, comparisons of these quantities between the two probes provide a useful measure of their relative atmospheric drag histories without requiring an absolute knowledge of the spacecraft ballistic coefficients. The integrals were evaluated over the portion of each orbital revolution below 600 km altitude. This upper limit encompasses the altitude range in which atmospheric drag becomes increasingly important while providing a common integration domain for both spacecraft. 

\subsection{Atmospheric Density at Common Altitudes}

To distinguish differences in atmospheric sampling from those arising directly from orbital altitude, the neutral densities encountered by the two spacecraft were compared at common reference altitudes. Atmospheric density was extracted within narrow altitude intervals ($\pm$5 km) centered at 200, 220, 250, and 300 km. Restricting the comparison to common altitude ranges substantially reduces the direct effect of the strong vertical density gradient and allows differences associated with spacecraft position and local time to be examined.

%\subsection{Orbital Geometry and Thermospheric Sampling}

%Because both the probes were exposed to the same large-scale solar and geomagnetic forcing, differences in their orbital geometry provide a potential mechanism for differential thermospheric sampling. To investigate this possibility, the evolution of the orbital elements and the geographic and local-time coordinates of perigee were examined for both spacecraft. They provide a first-order description of the thermospheric environment sampled during the portion of the orbit where atmospheric drag is strongest.
%
%The evolution of the orbital-plane geometry was additionally characterized using the right ascension of the ascending node $\Omega$, eccentricity $e$ and argument of perigee $\omega$. Differences between the spacecraft were evaluated using
%
%\begin{equation}
%	\Delta X = X_A-X_B,
%\end{equation}
%where $X$ represents the corresponding orbital parameter. Changes in these quantities can alter the altitude distribution and geographic location and local time of perigee, thereby modifying the thermospheric environment sampled by each spacecraft.

\subsection{Model Uncertainties}

NRLMSIS 2.0 represents an empirical estimate of the mean thermospheric state and therefore does not reproduce all short-term and spatially localized density variability. As a result, the magnitude of the modeled drag exposure is subject to uncertainties in the atmospheric model as well as in the orbital propagation. The present analysis therefore emphasizes relative differences between the two spacecraft evaluated with the same atmospheric model and forcing, rather than treating the modeled density or integrated drag as an exact measurement of the instantaneous atmospheric state. 

Because the orbital propagation is based on publicly available TLEs, uncertainties in the derived state vectors and orbital elements are expected, particularly during periods of rapid terminal decay. To minimize propagation error, each TLE was propagated only to the epoch of the next available TLE. The analysis focuses primarily on persistent differences that remain coherent over many orbital revolutions rather than on individual short-timescale fluctuations.

\section{Results}

\subsection{Divergence in Orbital Evolution}

\begin{figure}[hbt!]
	\centering
	\includegraphics[width=0.75\textwidth]{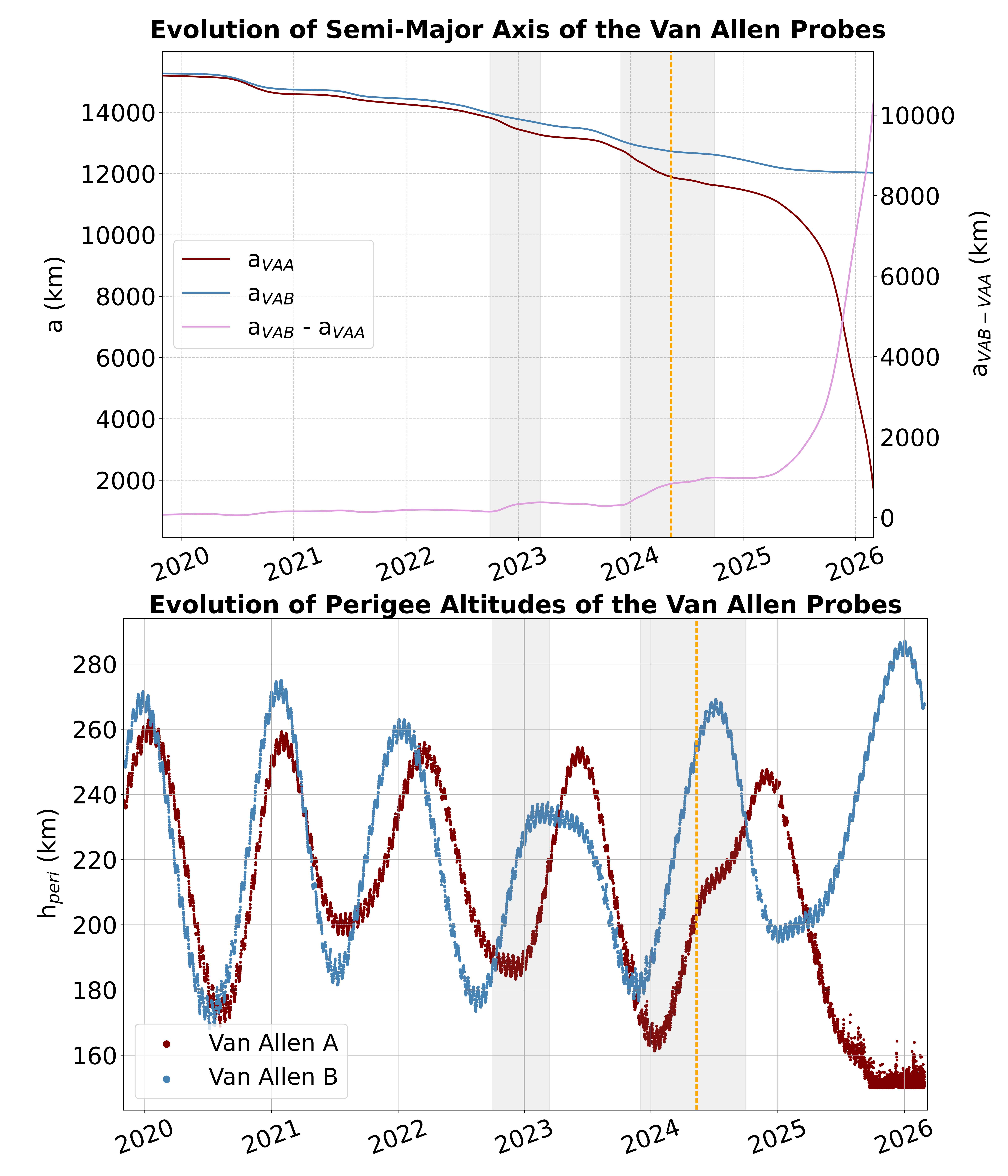}
	\caption{Evolution of the semi-major axes and perigee altitudes of the Van Allen Probes A and B following deactivation. The grey shaded regions highlight intervals of a pronounced increase in the difference between the semi-major axes of the two probes, as discussed further in Figure \ref{fig:DecayRates}. The orange lines indicate the occurrence of the Gannon storm between 10 and 13 May 2024.}
	\label{fig:decayprof_VanAnB}
\end{figure}

The top panel in Figure~\ref{fig:decayprof_VanAnB} shows the evolution of the semi-major axis of Van Allen Probes A and B following their deactivation in 2019. During the initial post-deactivation period, the two spacecraft exhibit broadly similar orbital evolution, with both semi-major axes decreasing gradually through the end of 2022. Superimposed on this common decay are shorter-timescale variations, but the long-term separation between the two spacecraft remains comparatively small. A systematic divergence becomes apparent during late 2022. The semi-major axis of Van Allen Probe A subsequently decreases more rapidly than that of Probe B, and the difference grows progressively till 2025. The divergence becomes particularly pronounced during the final stage of Probe A's evolution, when its semi-major axis decreases rapidly toward atmospheric reentry, whereas Probe B continues to evolve on a substantially higher orbit.

The bottom panel shows the evolution of the perigee altitudes of the two probes. In 2020, the perigee altitudes and the corresponding perigee envelopes, which provide an indication of the residence time at low altitudes, are comparable for both probes, with the lowest altitudes of both envelopes reaching below 170 km. Taking 220 km as a reference altitude, both probes spend approximately the same amount of time below this altitude during this period.

In 2021, however, the perigee evolution of the two probes begins to diverge. During the first oscillation, Van Allen Probe B reaches a minimum perigee altitude of approximately 180 km, whereas Van Allen Probe A reaches approximately 200 km. Although Probe A does not reach as low a minimum altitude, its perigee envelope is broader. These small differences in orbital evolution subsequently become amplified through differential atmospheric drag, as the lower-altitude portions of the orbit experience substantially greater thermospheric density. During the subsequent oscillation, Probe B reaches a minimum altitude of approximately 175 km, while Probe A reaches approximately 185 km. Again, the perigee envelope of Probe A is broader and exhibits a further forward offset relative to that of Probe B.

\begin{figure}[hbt!]
	\centering
	\includegraphics[width=0.9\textwidth]{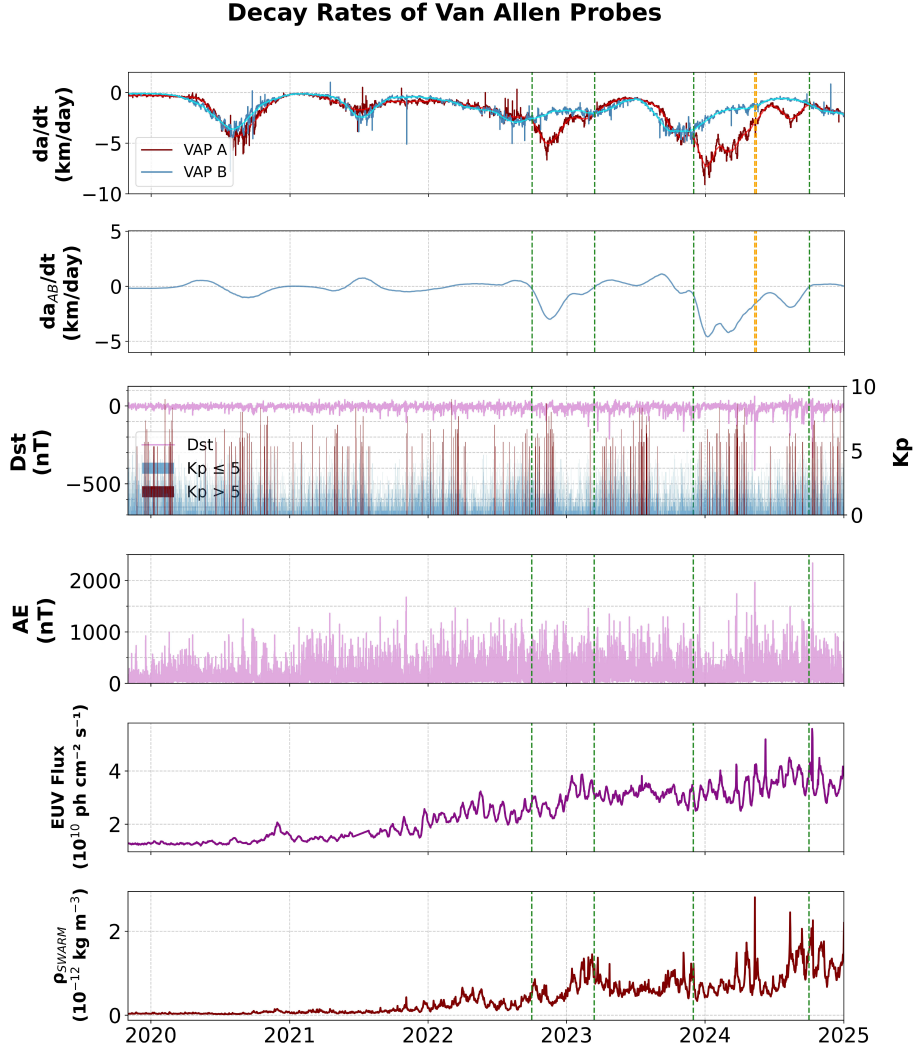}
	\caption{Temporal evolution of the semi-major axis decay rates of Van Allen Probes A and B, difference in their 27-day averaged decay rates, together with the geomagnetic $Dst$/Kp and $AE$ indices, solar EUV flux and daily mean atmospheric densities measured by SWARM-B at the corresponding altitude. The green vertical lines mark the reference epochs $d_1=$ 1 October 2022, $d_2=$ 15 March 2023, $d_3=$ 1 December 2023, and $d_4=$ 1 October 2024. The orange lines in the top two panels indicate the occurrence of the Gannon storm between 10 and 13 May 2024. The EUV flux is derived from measurements by the Solar Extreme Ultraviolet Monitor (SEM) aboard SOHO.}
	\label{fig:DecayRates}
\end{figure}

The grey regions indicate periods during which there is a sudden increase in the difference in the semi-major axes of the two probes, as emphasized further in Figure \ref{fig:DecayRates}. Following the first such deviation, around 1 October 2022, the perigee altitude of Probe B remains above approximately 190 km, while that of Probe A falls below 190 km. During the subsequent oscillation of the perigee envelopes, extending from mid-2023 to 2025, Probe A reaches a substantially lower minimum perigee altitude, near 160 km, compared with approximately 180 km for Probe B. The offset between the two perigee envelopes is also considerably larger during this period, with Probe A exhibiting a broader envelope.

Around mid-2024, possibly in association with the Gannon storm in May 2024, the recovery of the perigee altitude of Probe A appears to be further slowed, whereas the perigee evolution of Probe B remains comparatively unaffected. This difference warrants further investigation of the geographical locations and local times of the two probes during the relevant storm periods and at the corresponding altitudes. In addition, the physical triggers responsible for the first two deviations in the orbital evolution need to be identified.

The temporal evolution of the semi-major axis decay rates, together with the geomagnetic $Dst$ and $AE$ indices and solar EUV flux, is shown in Figure~\ref{fig:DecayRates}. The decay rates of the two spacecraft remain broadly similar during the early post-deactivation period, although short-lived differences do occur throughout. A more persistent separation begins to develop during 2022-2023 and becomes pronounced from late 2023 onward. During 2024, the decay rate of Probe A becomes systematically more enhanced compared to Probe B, with the largest differences occurring during the final phase of Probe A's orbital evolution.

\begin{figure}[hbt!]
	\centering
	\includegraphics[width=\textwidth]{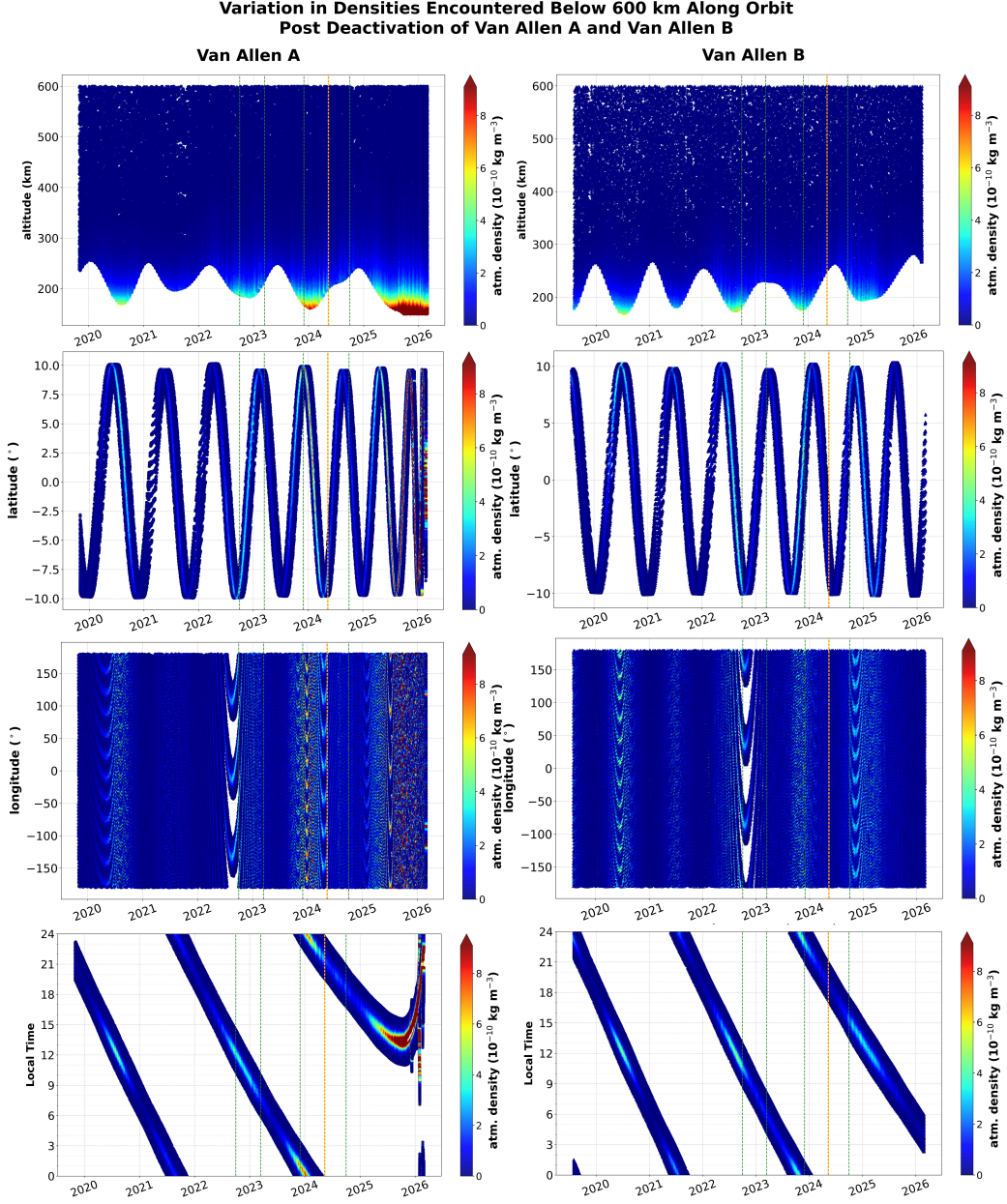}
	\caption{Orbital evolution (below 600 km) and model thermospheric densities of Van Allen Probes A and B following deactivation. The left and right columns show, respectively, the altitude, geographic latitude, geographic longitude, and local solar time of Probes A and B; the points are colored by NRLMSIS 2.0 neutral density.}
	\label{fig:Den_AltnLatvsEpoch}
\end{figure}

Figure~\ref{fig:Den_AltnLatvsEpoch} shows the altitude, latitude, longitude, and local time sampled along the satellite orbits at altitudes below 600 km, with the color indicating the atmospheric density obtained from the NRLMSIS model. Notably, the perigee evolution of Probe B changes character during 2025. Unlike the preceding oscillations, which repeatedly reach perigee altitudes near 180-200 km, the final oscillation remains substantially higher, with the perigee envelope not dropping below 200 km. This occurs despite the continued decrease in semi-major axis. 

During the Gannon storm, Probe A remains at local times between 19 and 24 LT, whereas Probe B remains between 17 and 21 LT. This difference in local-time sampling may provide a possible explanation for the contrasting orbital responses of the two probes during the storm, and may point to a local-time dependence of the thermospheric density enhancement, with a potentially stronger enhancement during the late-evening sector between 21 and 24 LT.

%The evolution of the orbital-plane geometry is shown in Figure \ref{fig:OrbitalParDiff}.

%\begin{figure}[hbt!]
%	\centering
%	\includegraphics[width=0.85\textwidth]{OrbitalParDiff.jpg}
%	\caption{Evolution of differences in semi-major axes, argument of perigees, eccentricities and right ascension of ascending nodes between Van Allen Probes A and B post decommissioning.}
%	\label{fig:OrbitalParDiff}
%\end{figure}

\subsection{Divergence in Thermospheric Sampling}

\begin{figure}[hbt!]
	\centering
	\includegraphics[width=1.0\textwidth]{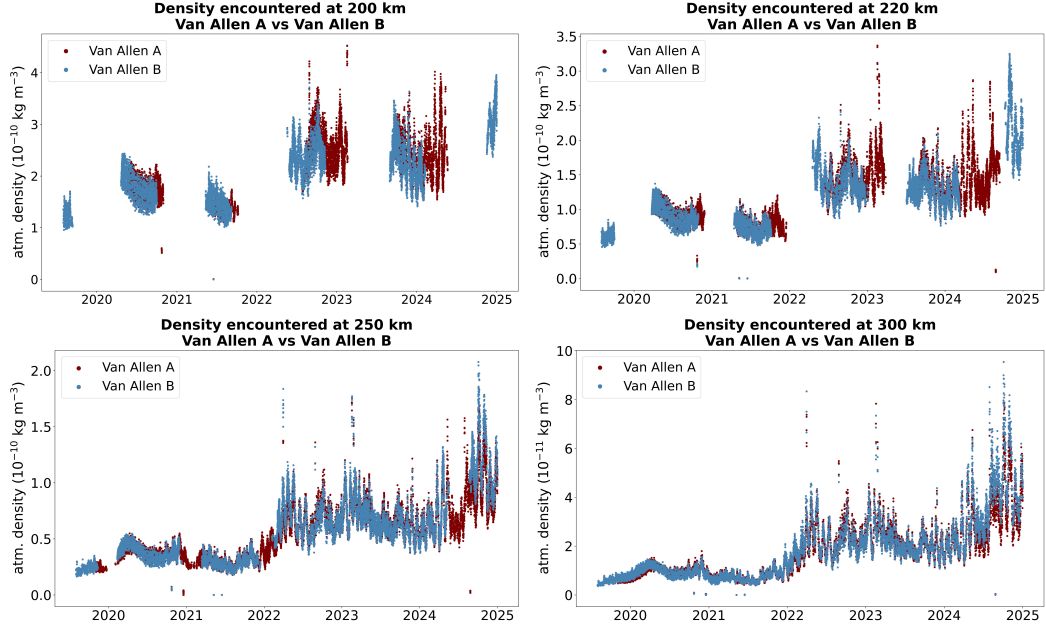}
	\caption{Atmospheric densities encountered by Van Allen Probes A and B following deactivation, till 2025, at altitudes of 200$\pm$5 km, 220$\pm$5 km, 250$\pm$5 km, and 300$\pm$5 km.}
	\label{fig:Den_SameAlts_VanAvsB}
\end{figure}

The divergence in orbital evolution raises the question of whether the two spacecraft also sampled systematically different thermospheric densities. Because atmospheric density varies strongly with altitude, a direct comparison of the densities encountered along the complete trajectories cannot distinguish differences caused by orbital altitude from differences associated with geographic and temporal sampling. We therefore compared the modeled neutral densities encountered by the two spacecraft following deactivation, till 2025, within narrow altitude bands centered at 200, 220, 250, and 300 km (Figure~\ref{fig:Den_SameAlts_VanAvsB}). The analysis is limited to 2025 to isolate the period over which the divergence in orbital evolution developed, before the subsequent evolution became substantially different.

During the early post-deactivation period, the density distributions within the same altitude bands are broadly similar for the two spacecraft. The agreement is particularly evident at 250 and 300 km, where the density time series exhibit similar long-term variations until late 2022. Increasing differences emerge thereafter, becoming more apparent at 250 km and most pronounced at 220 and 200 km. This altitude dependence suggests that the increasing difference in atmospheric exposure cannot be attributed solely to Probe A reaching lower altitudes, since systematic differences in the modeled densities are also observed when the two spacecraft occupy comparable altitude ranges. However, the temporal sampling of these altitude bands must also be considered. From late 2022 onward, the perigee envelope of Probe A becomes progressively broader, allowing it to continue sampling the 200 and 220 km altitude ranges for several months longer than Probe B. Consequently, even at the same altitude, the two probes may sample these regions at different epochs and therefore under different thermospheric conditions.

To examine whether differences in orbital sampling could further contribute to this behavior, the geographic longitude and local time sampled by the two spacecraft were compared (Figure~\ref{fig:Den_AltnLatvsEpoch}). Both spacecraft exhibit systematic long-term evolution of their perigee location following deactivation, reflecting the precession of their orbital geometry. The two trajectories remain broadly similar during the early post-deactivation period but progressively develop differences in their longitude and local time sampling as the orbital evolution proceeds. As a result, common solar and geomagnetic forcing does not imply identical thermospheric sampling (\citet{Tulasi_Ram_2016,Wei_2024}): differences in orbital geometry can cause the two spacecraft to encounter different combinations of latitude, longitude, and local solar time. Such differences provide a possible pathway for generating the density differences observed at the same altitude bands.

There are two effects in play here. The first is the direct effect of orbital altitude, where Probe A progressively enters regions of higher neutral density. The second is a sampling effect, whereby the two spacecraft can encounter different atmospheric states even at comparable altitude because they occupy different locations and local time sectors. Both effects become increasingly important during the later evolution of Probe A and are examined together with the low-altitude residence time and integrated drag in the following section.

\subsection{Differential Low-Altitude Residence and Drag Exposure}

The same-altitude density comparison demonstrates that the two spacecraft increasingly sampled different thermospheric conditions during the later stages of their orbital evolution. However, the cumulative effect of these differences depends not only on the density encountered but also on the duration for which each spacecraft remains within the dense lower thermosphere. The residence time below a range of critical altitudes and integrated atmospheric drag accumulated during each orbital revolution were examined for both the probes. For each spacecraft, the time spent below a specified altitude $h_c$ was calculated for every orbital revolution. This measures the duration of exposure to the lower thermosphere without imposing a particular atmospheric density model.

\begin{figure}[hbt!]
	\centering
	\includegraphics[width=0.75\textwidth]{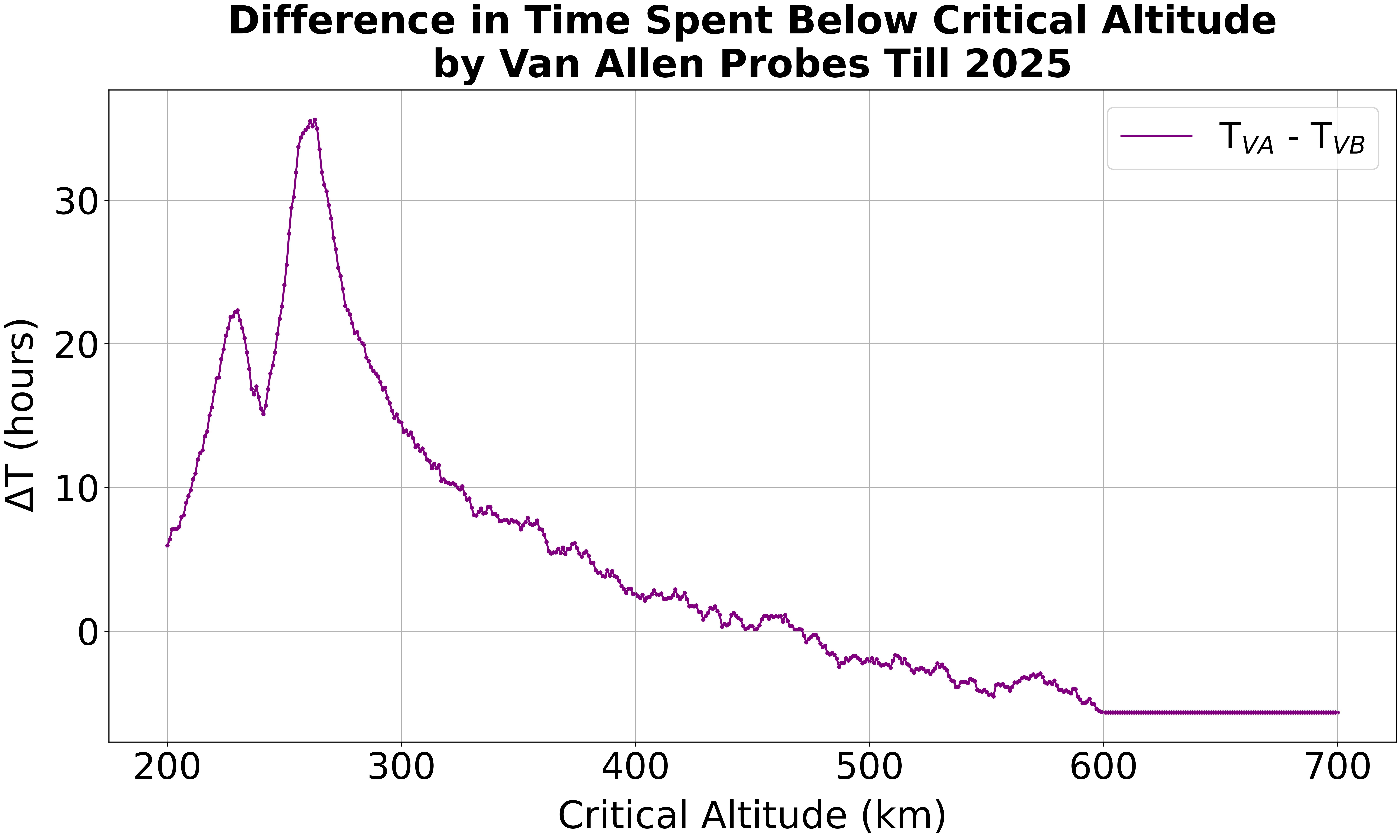}
	\caption{Difference in cumulative residence time below a critical altitude between Van Allen Probes A and B. The maximum separation occurs near 263 km, where Van Allen Probe A accumulates approximately 39 hours more residence time than Van Allen Probe B over the analyzed period.}
	\label{fig:TimeDiffBelowAlt}
\end{figure}

The cumulative difference in residence time till 2025
\begin{equation}
	\Delta T_{<h_c}
	=
	T_{A,<h_c}-T_{B,<h_c},
\end{equation}
was evaluated for altitudes between 200 and 700 km. As shown in Figure~\ref{fig:TimeDiffBelowAlt}, the largest positive difference occurs near 263 km, where Probe A accumulates around 39 hours of extra residence time below the critical altitude compared to Probe B over the same period.

Figure~\ref{fig:TimeDiffBelowAlt} shows that Probe A accumulates substantially greater cumulative residence time than Probe B at low critical altitudes, with the maximum difference of approximately 39 hours occurring near 263 km. The difference then decreases progressively with increasing critical altitude, approaching zero near 470 km as Probe B accumulates relatively more residence time at higher altitudes. Above around 470 km, Probe B has the greater cumulative residence time.

\begin{figure}[hbt!]
	\centering
	\includegraphics[width=0.7\textwidth]{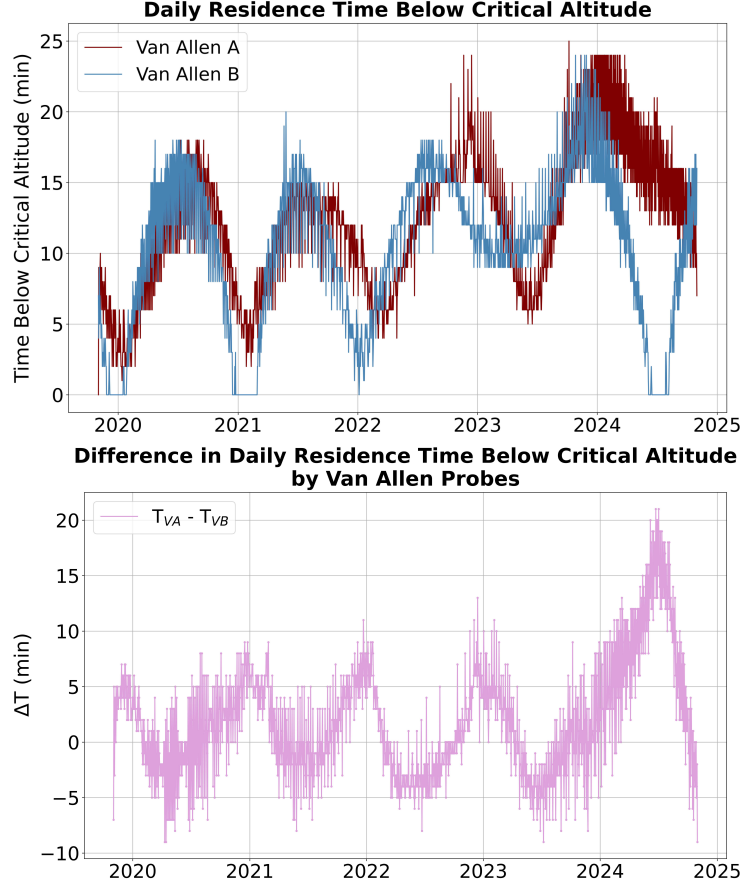}
	\caption{Daily residence time below 263 km for Van Allen Probes A and B and their difference following deactivation. The increasing positive difference during 2024 indicates that Van Allen Probe A progressively spent more time within the lower thermosphere than Van Allen Probe B.}
	\label{fig:DailyTimeBelow}
\end{figure}

The evolution of daily residence times below this critical altitude for both spacecraft is shown in Figure~\ref{fig:DailyTimeBelow}. During the early post-deactivation period, the daily residence times of the two spacecraft exhibit broadly similar variations, and neither spacecraft maintains a persistent advantage in low-altitude exposure. During 2023, the daily residence-time difference becomes increasingly asymmetric, and during 2024, Probe A increasingly spends more time below the critical altitude than Probe B. 

The residence time difference alone does not determine the difference in atmospheric drag. Because the neutral density increases rapidly with decreasing altitude, the drag contribution from a given interval of time depends strongly on the altitude and thermospheric state during that interval. We therefore calculated the integrated drag exposure for each orbital revolution using equation 1. The corresponding specific orbital energy loss expected from atmospheric drag is proportional to
\begin{equation}
	E_D
	=
	\frac{1}{2}
	\frac{C_D A}{m}
	\int_{\mathrm{orbit}}
	\rho v^{3}\,dt.
\end{equation}
The quantities $\int\rho v^2dt$ and $\int\rho v^3dt$ are therefore used as spacecraft-independent measures of cumulative atmospheric drag exposure and drag-related energy dissipation, respectively.

The increasing drag exposure should be reflected in the secular loss of orbital energy. For a Keplerian orbit, the specific orbital energy is

\begin{equation}
	E_{\mathrm{orb}}=-\frac{\mu}{2a},
\end{equation}

where $\mu$ is the Earth's gravitational parameter and $a$ is the semi-major axis. The change in specific orbital energy between successive orbital revolutions was calculated as

\begin{equation}
	\Delta E_{\mathrm{orb},i}
	=
	E_{\mathrm{orb},i+1}-E_{\mathrm{orb},i}.
\end{equation}

Because atmospheric drag removes orbital energy, negative values of $\Delta E_{\mathrm{orb}}$ correspond to orbital energy loss.

\begin{figure}[hbt!]
	\centering
	\includegraphics[width=0.7\textwidth]{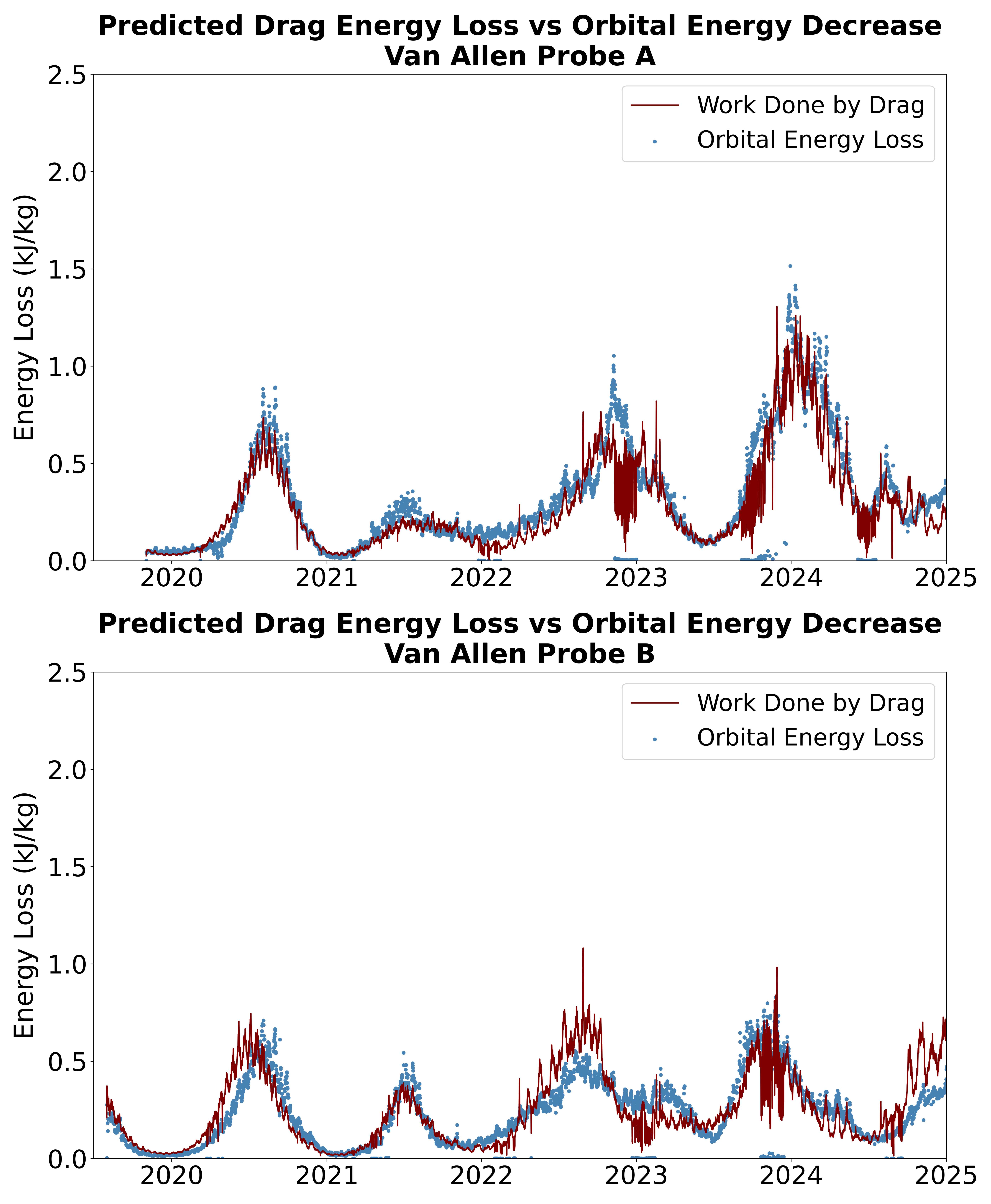}
	\caption{Comparison of predicted drag-induced energy loss and orbital energy loss experienced by the Van Allen probes.}
	\label{fig:EnergyLossComp}
\end{figure}

Figure~\ref{fig:EnergyLossComp} compares the orbital energy loss computed from the evolution of the semi-major axis with the corresponding drag-derived energy loss estimate for the two spacecraft. The two quantities exhibit broadly similar temporal behavior over much of the post-deactivation period. Periods of enhanced drag energy dissipation are generally accompanied by enhanced orbital energy loss.

To compare the magnitudes of the two quantities, an effective drag coefficient of $C_D=2.2$ was adopted together with an average spacecraft area estimate of 99.526 m$^2$ available in the DISCOS database. The drag-derived energy loss required scaling by 1/10 to match the magnitude of the observed orbital energy decrease. This factor is treated here as an empirical normalization rather than as a modification of the physical drag relation. The difference in magnitude may arise from several sources of uncertainty, including the effective velocity-facing area, spacecraft attitude, the assumed drag coefficient, and uncertainties in the empirical atmospheric density model. The probes were reported to have a mean projected spacecraft area of $5.0421~\mathrm{m^2}$ and a median of $5.0163~\mathrm{m^2}$ based on the distribution of velocity-facing orientations \citep{Bull2020}; their analysis also emphasized the sensitivity of orbit prediction to uncertainties in the effective area and drag coefficient. In addition, the present analysis accounts for atmospheric drag only below 600 km, where the density is substantially higher, and drag effects are expected to be most significant. Excluding drag contributions above this altitude introduces an additional source of discrepancy in the absolute energy-loss estimate. Consequently, the comparison is used primarily to evaluate the temporal consistency between atmospheric drag and orbital energy evolution rather than their absolute magnitudes.

During 2019-2022, the two spacecraft exhibit broadly comparable drag exposure, although individual intervals of enhanced atmospheric density produce short-timescale variability. As the orbital evolution diverges, a persistent separation develops in the drag-derived energy loss, with Probe A increasingly experiencing greater energy dissipation than Probe B. This difference becomes particularly pronounced during 2024 and 2025, when Probe A enters a rapidly accelerating phase of orbital decay.

The enhanced drag experienced by Probe A is consistent with its increasing residence time at low altitude. Because the atmospheric density gradient becomes increasingly steep toward lower altitudes, additional residence time in this region contributes disproportionately to the total drag exposure. The relationship between residence time and drag is therefore nonlinear: a small downward change in perigee does not simply add a comparable amount of drag exposure, but places the spacecraft in an environment where the neutral density can increase substantially over a relatively small change in altitude. Once Probe A begins to spend a larger fraction of each revolution in this region, the resulting increase in drag enhances its orbital energy loss and further lowers its orbit.

The residence-time, density, and energy-loss analyses therefore indicate an amplifying feedback between orbital evolution and atmospheric interaction. The feedback is not established by residence time alone, but by the combination of increased low-altitude residence, the steep vertical density gradient, and the evolving thermospheric conditions sampled by the spacecraft. This provides a pathway by which relatively small differences in thermospheric exposure can develop into increasingly different orbital energy losses and, ultimately, the substantially divergent orbital evolution of the two probes.

\subsection{Differential Orbital Decay and Space Weather Forcing}

The earlier sections show that the divergence between the two spacecraft is accompanied by increasing differences in low-altitude residence, same-altitude thermospheric density, integrated drag exposure, and orbital energy loss. Because both probes were exposed to the same large-scale solar and geomagnetic environment, the next question is whether changes in external forcing can account for the onset of their differential orbital decay.

%To further investigate the sudden differences in decay rates, the intervals between $d_1$ and $d_2$, and between $d_3$ and $d_4$, are examined in detail (Figure \ref{fig:DecayRates_d1d2d3d4}). 

Storms with minimum Dst values between -30 and -50 nT are classified as weak storms, while those with minimum Dst values between -50 and -100 nT are classified as moderate storms. Storms with minimum Dst values between -100 and -200 nT are considered strong storms, whereas those between -200 and -350 nT are classified as severe storms. Finally, storms with minimum Dst values below -350 nT are classified as intense or great storms \citep{ColladoVillaverde2024}.

Between $d_1$ and $d_2$, a G1 class minor geomagnetic storm occurred in October 2022. A weak -40 nT storm occurred on 2 October, which was followed by a relatively slow recovery. An SSC and a subsequent moderate-intensity -60 nT storm occurred on 14 October, exhibiting a faster recovery. This was followed by another moderate-intensity storm of -80 nT on 22 October. The maximum difference in decay rates occurs during the recovery phase of this latter storm.

During the February 2022 G1-class geomagnetic storm, associated with the loss of 38 Starlink satellites, density enhancements near 210 km peaked during the recovery phase, about 8-10 h after the Dst minimum \citep{Berger2022}. Moderate geomagnetic storms can therefore produce substantial density enhancements at the altitudes relevant to the Van Allen Probes. In fact, \citet{Berger2022} reported enhancements of approximately 20-30\% near 200 km during the February 2022 storm.

By the end of October, Probe A had reached a perigee altitude of approximately 180 km, while Probe B remained above 200 km. Thus, during the recovery phase of the moderate-intensity storm that occurred on 22 October, the two spacecraft were sampling different portions of the lower thermosphere. Given the steep increase in density with decreasing altitude, the same fractional storm-time enhancement would produce a substantially larger drag acceleration on Probe A. Moreover, the storm-time density response is spatially structured, with enhancements driven by Joule heating over the polar region and subsequent thermospheric expansion to equatorial latitudes \citep{Kataoka2022}. Differences in altitude and orbital location could therefore lead to markedly different drag responses.

The October 2022 storm may consequently have provided a perturbation that amplified the pre-existing orbital differences between the probes. The additional drag on the lower-altitude Probe A would cause further orbital energy loss and perigee reduction, increasing its subsequent exposure to the denser lower thermosphere. This provides a possible positive feedback mechanism linking the storm-time thermospheric response to the subsequent divergence in orbital decay.
	
The second major deviation in the decay rates between the two probes occurs following several strong storms with Dst reaching approximately -100 nT in December 2023. The 1 December event was reported to trigger unusual equatorial plasma bubbles (EPBs) over longitudes of 100$^\circ$-140$^\circ$E, which persisted for several hours after sunrise on the following day \citep{Rajesh2025}. At the time of the maximum difference in decay rates, the AE index also increased abruptly to approximately 1500 nT from an average level of around 400 nT, indicating enhanced geomagnetic energy input. Another strong storm, reaching approximately -100 nT, occurred on 2 March 2024. Following the 24 March 2024 storm (Dst$\sim$-130 nT), the neutral atmospheric temperature at 250 km increased from approximately 1000 K to 1500 K within a few hours \citep{KHABITUEV2026}. The next major deviation occurs during the G4-class storm on 12 August 2024 (Dst$\sim$-190 nT), which was preceded by another strong storm reaching approximately -100 nT on 4 August.

The evolution of the solar EUV flux provides evidence for a substantial change in the background thermospheric forcing during the same period. EUV flux increases progressively from the relatively low solar-activity conditions of 2019 toward the rising phase of Solar Cycle 25. Enhanced solar EUV heating increases thermospheric temperature and, consequently, neutral density at satellite altitudes. The increasing EUV flux therefore provides a plausible explanation for the overall increase in atmospheric drag observed during the later post-deactivation period.

The geomagnetic indices exhibit shorter-timescale variability superimposed on this longer-term solar cycle evolution. Episodes of $Dst$ minima and enhanced $AE$ indicate periods of increased geomagnetic forcing, during which thermospheric density can be substantially enhanced. Several such intervals occurred during 2023-2024, coincident with periods of enhanced orbital decay. However, these disturbances affect both spacecraft simultaneously and therefore cannot, by themselves, explain the persistent difference between their long-term decay rates.

The distinction between common forcing and differential response is evident in Figure~\ref{fig:DecayRates}. The green lines correspond to $d_1=$ 1 October 2022, $d_2=$ 15 March 2023, $d_3=$ 1 December 2023, and $d_4=$ 1 October 2024. Between $d_1$ and $d_2$ as well as between $d_3$ and $d_4$, there is a significant visible enhancement in the decay rate of Van Allen A over Van Allen B. During intervals of enhanced solar and geomagnetic activity, both spacecraft can exhibit increased orbital decay, but the magnitude of the response becomes progressively different as their orbital evolution separates. In particular, the persistent excess decay of Probe A extends over timescales substantially longer than individual geomagnetic disturbances. The temporal agreement between the onset of the persistent decay-rate difference and the sequence of geomagnetic disturbances in October 2022 suggests that this period may have provided an initial trigger. However, because both spacecraft experienced the same external forcing, the differential response must have depended on spacecraft-specific orbital or physical characteristics. Once established, the initial difference appears to have been amplified by the subsequent divergence in perigee altitude, low-altitude residence time, and atmospheric drag exposure.

%Although subsequent major storms-such as those in April 2023 and October 2024-occurred after the Van Allen Probes' deactivation, they impacted both spacecraft similarly without causing further differential decay..

The increasing EUV flux may nevertheless have played an important indirect role. As the background thermospheric density increased during the rising phase of Solar Cycle 25, the drag penalty associated with a given difference in altitude or atmospheric sampling also increased. Thus, a relatively small pre-existing difference in orbital geometry could produce a progressively larger difference in atmospheric drag under stronger solar forcing. In this sense, solar activity may have acted as an amplifier of spacecraft-specific orbital differences rather than as a unique driver of the divergence.

The temporal evolution of the decay rates is consistent with this interpretation. Differences between the spacecraft become detectable before the strongest separation in 2024, while the largest differences in integrated drag exposure, low-altitude residence time, and orbital energy loss occur later. The sequence is therefore more consistent with a gradually amplified differential response than with a single impulsive space weather event initiating the divergence.

The simultaneous evolution of solar activity and orbital decay therefore provides important environmental context but does not, by itself, identify the initiating mechanism. Both spacecraft experienced the same external solar and geomagnetic forcing, whereas their orbital geometries and atmospheric sampling evolved differently. The observations are consequently most consistent with an interaction between common thermospheric forcing and spacecraft-specific orbital evolution: increasing solar activity raised the background density and strengthened the drag sensitivity of the orbits, while small differences in orbital sampling produced different cumulative responses that were subsequently amplified through orbital decay.

A key implication is that the divergence cannot be attributed to the occurrence of a particular geomagnetic storm without accounting for the pre-existing differences in orbital geometry. Instead, the observations support a scenario in which slowly varying solar forcing establishes the background thermospheric state, transient geomagnetic activity modulates the density, and differences in orbital sampling determine how strongly each spacecraft responds. The resulting differential energy loss can then alter the orbital trajectories themselves, providing a feedback that progressively increases the contrast between the two spacecraft.

\section{Discussion}

The substantially different orbital lifetimes of Van Allen Probes A and B cannot be explained solely by their initial orbital conditions or by the common solar and geomagnetic forcing experienced after deactivation. The two spacecraft initially followed broadly similar orbital evolution, but small differences in their perigee envelopes and orbital geometry emerged from 2021 onward and became substantially amplified after late 2022. Probe A subsequently developed lower and broader perigee envelopes, increasing its residence time in the lower thermosphere and its cumulative atmospheric drag exposure. This suggests that the observed divergence developed through an amplifying feedback rather than from a single isolated event.

The model density comparison provides evidence that differences in atmospheric sampling contributed to this process. During the early post-deactivation period, the densities encountered by the two spacecraft within the same altitude bands are broadly similar, particularly at 250 and 300~km. Increasing differences appear after late 2022 and are strongest at 200 and 220~km. Although part of this difference results from Probe A sampling these low altitudes for longer periods as its perigee envelope broadens, differences in geographic longitude, latitude, and local time also develop between the two spacecraft. Thus, identical solar and geomagnetic forcing does not imply identical thermospheric exposure.

The first major change in the decay-rate difference occurs around October 2022 and is particularly noteworthy. Several storms occurred during this period, with $Dst$ reaching approximately $-40$, $-60$, and $-80$~nT. The maximum difference in decay rates occurs during the recovery phase of the 22-23 October storm. This timing is consistent with the delayed thermospheric response observed during the February 2022 storms associated with the loss of 38 Starlink satellites, for which density enhancements near 200-210~km persisted into the recovery phase \citep{Berger2022}. Kataoka et al.~\citep{Kataoka2022} further showed that moderate geomagnetic storms can produce density enhancements of up to approximately 50\% at 200~km.

This is particularly relevant because, by the end of October 2022, Probe A had reached a perigee altitude of approximately 180~km, whereas Probe B remained above 200~km. A storm-time density enhancement would therefore produce a substantially larger drag acceleration on Probe A because of both its lower altitude and potentially different geographic and local-time sampling. The October storm sequence may consequently have acted as an important amplifying perturbation: additional energy loss by Probe A would lower its perigee further, increasing its subsequent exposure to the much denser lower thermosphere and establishing a positive feedback between altitude and drag.

The later evolution suggests that this process was reinforced by subsequent geomagnetic disturbances. A second major deviation in decay rates occurs following several strong storms in December 2023, including an abrupt increase in $AE$ to approximately 1500~nT. The 1 December event was also associated with unusual equatorial plasma bubbles over $100^\circ$--$140^\circ$E that persisted into the post-sunrise period \citep{Rajesh2025}, indicating a strongly disturbed upper-atmospheric state. Further strong storms occurred in March 2024, including the 24 March event ($Dst\sim-130$~nT), following which the neutral atmospheric temperature at 250~km reportedly increased from approximately 1000 to 1500~K within several hours \citep{KHABITUEV2026}. The next major deviation occurs during the G4 storm of 12 August 2024 ($Dst\sim-190$~nT), preceded by another strong storm (Dst$\sim-100$~nT) on 4 August. These events suggest that geomagnetic activity repeatedly amplified the difference in decay rates as the orbital configurations of the two spacecraft became increasingly different.

The overall evolution therefore appears to consist of a gradual initial orbital divergence followed by episodic storm-time amplification and an increasingly strong altitude-drag feedback. The increasing solar EUV flux during Solar Cycle~25 provides a slowly varying background enhancement of thermospheric density, while geomagnetic storms produce transient disturbances. Once Probe A entered a lower-perigee regime, these disturbances could produce a disproportionately larger drag response on Probe A than on Probe B. The resulting additional orbital energy loss further lowered Probe A's perigee, increasing its exposure to atmospheric drag. 

This interpretation also explains the increasing separation in residence time, integrated $\int\rho v^2dt$ drag exposure, and $\int\rho v^3dt$ energy-loss proxies. By 2024-2025, Probe A had accumulated substantially greater low-altitude exposure, and the resulting feedback had become sufficiently strong to produce its rapidly accelerating terminal decay.

The MSIS 2.0 densities should nevertheless be interpreted cautiously. As demonstrated by Kataoka et al.~\citep{Kataoka2022}, empirical models can underestimate the magnitude and spatial extent of storm-time thermospheric disturbances. The modeled densities therefore provide evidence for relative differences in atmospheric sampling rather than direct measurements of the actual storm-time density. 

Overall, the results suggest that the different lifetimes of the two probes resulted from convolution of the interaction of small initial orbital differences, episodic geomagnetic-storm amplification, and nonlinear altitude-dependent atmospheric drag, rather than from any single factor acting independently.

\section{Conclusion}

The divergent orbital lifetimes of the Van Allen Probes resulted from the combination of   progressive interaction of orbital evolution, thermospheric drag, and space weather variability. Small differences in their orbital evolution developed after deactivation and were amplified as Probe A entered lower perigee altitudes, increasing its residence time and exposure to the dense lower thermosphere. Geomagnetic disturbances further enhanced this difference, leading to progressively greater drag and orbital energy loss for Probe A and ultimately to its reentry in March 2026, while Probe B remained in orbit. These results demonstrate how small differences between spacecraft sharing nearly identical initial orbits can be amplified through the nonlinear response of atmospheric drag, producing substantially different orbital lifetimes under the same external forcing.

\section*{Open Research Section}

All data used in this study are publicly available. TLE data were obtained from Space-Track (space-track.org). The Dst and AE geomagnetic indices were acquired from the WDC, Kyoto website (wdc.kugi.kyoto-u.ac.jp). The $F_{10.7}$ and $Ap$ solar and geomagnetic indices were taken GFZ, Potsdam website (kp.gfz-potsdam.de/en/data).

\section*{Conflict of Interest declaration}
The authors declare there are no conflicts of interest for this manuscript.

\bibliography{agusample}

\end{document}